\documentclass[aps,prl,twocolumn,showpacs,amsmath,amssymb,groupedaddress]{revtex4}

\usepackage{graphicx}% Include figure files
\usepackage{dcolumn}% Align table columns on decimal point
\usepackage{bm}% bold math
\begin{document}

\preprint{APS/123-QED}

\title{A Smooth, Inductively Coupled Ring Trap for Atoms}

\author{P. F. Griffin}
\author{E. Riis}
\author{A. S. Arnold}
\affiliation{Department of Physics, SUPA, University of Strathclyde, Glasgow G4 0NG, UK}
\date{\today}

\begin{abstract}
We propose and numerically investigate a scalable ring trap for cold atoms that surmounts problems of roughness
of the potential and end--effects of trap wires. A stable trapping potential is formed about an electrically
isolated, conducting loop in an ac magnetic field by time averaging the superposition of the external and
induced magnetic fields. We investigate the use of additional fields to eliminate Majorana spin flip losses and
to create novel trapping geometries. The possibility of micro--fabrication of these ring traps offers the
prospect of developing Sagnac atom interferometry in atom--chip devices.

\end{abstract}

   % Ultracold gases, trapped gases

\maketitle

There is significant interest amongst the cold atom community to build devices utilizing the sensitivity of
matter--wave optics. This research is driven not only by the desire to explore new regimes of atomic physics,
but also by the possibility of building sensors based on matter--wave interferometry
\cite{HindsBoshier:PRL:01,Schmiedmayer:PRL:02}. An extremely promising method of achieving these goals is
through micro--fabricated structures for trapping and manipulating atoms on atom chips
\cite{Folman:review:02,Zimmerman:RevModPhys:07}. Trapping and manipulation of atoms near permanent magnetic
structures has been an active research topic for a number of years \cite{Hinds:JPhysD:99}, much current work is
focussed on electromagnetic trapping, as this allows dynamic control of the trapping potentials. With relatively
modest currents, on the order of 1~A, large magnetic field gradients and curvatures can be obtained due to the
close proximity of the atoms to the wires creating the magnetic potentials. Furthermore, due to the use of
modern semiconductor fabrication techniques, complex structures can be constructed \cite{Schmiedmayer:xxx:08}.
One example of this was the demonstration of the controlled transport of atomic clouds along an atom chip
`conveyor belt' \cite{Haensch:PRL:01}.

A leading motivation behind atom chip experiments is the desire to explore reduced dimensional quantum gasses
\cite{Schmiedmayer:NatPhys:05}. However, fragmentation of atomic clouds close to trapping wires has been
observed by many groups \cite{Zimmerman:JPB:02,Pritchard:PRL:02,Pritchard:PRL:03,Hinds:PRL:03}, which presents a
significant problem for single--mode wave guiding of atoms and coherent matter--wave optics. In
\cite{Zimmerman:JPB:02,Pritchard:PRL:03} this roughness of the trapping potential was shown to be due to
longitudinal magnetic fields arising from deviations of the current flow from the desired path. By recognizing
that the roughness of the potential is proportional to the current through the trap wires, these defects were
recently circumvented through the use of ac currents to generate the magnetic fields \cite{Westbrook:PRL:07}.
With this method the corrugation of the trapping potential averages to zero while still maintaining the desired
trap depths and frequencies.

A complementary approach to explore reduced dimensional quantum gasses are ring traps. Even in the largest
aspect ratio 1D traps there is finite confinement and extension along the `weak' trapping direction. In a ring
trap the weak direction has infinite extent with periodic boundary conditions. This opens the way for
investigation of superfluid phenomena in quantum gasses \cite{McCann:PRA:03,Phillips:PRL:07}. For sensor
development a symmetric ring potential is ideal for a Sagnac atom interferometer
\cite{Chapman:PRL:01,Arnold:PRA:06} as the shared paths of opposing arms of the interferometer provide common
mode rejection of noise due to path--length differences. To date, magnetic ring traps have involved the use of
magnetic or electro--magnetic elements to create a multiply connected magnetic field minimum in circular
\cite{Chapman:PRL:01,Arnold:JPhysB:04,StamperKurn:PRL:05,Arnold:PRA:06} and stadium \cite{Prentiss:PRA:04}
geometries. In these cases electrical connectivity of the coils to an external power supply leads to a
perturbation of the ideal symmetric field. When the traps are scaled down, and the ring trap forms close to the
current carrying coils, these effects become increasingly significant.

\begin{figure}
\includegraphics[width=0.45\textwidth]{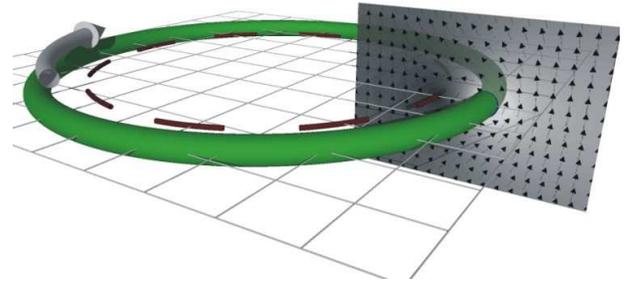}\\%addition_of_fields.eps}\\
\caption{\label{AdditionOfFields} Schematic of the instantaneous vector fields for the system described in the
text. The external bias field, increasing in magnitude, points upwards, while the induced current in the loop,
indicated by the arrow, creates a field about the loop. The fields cancel at a radius inside the loop, indicated
by the red, dashed circle. The grey--scale in the field slice indicates field magnitude and arrows the field
direction.} \vspace{-0.5cm}
\end{figure}

Here we propose a scalable, smooth ring trap for ultra--cold atoms. No external electrical connection is
required, eliminating undesirable end effects of wires and maintaining the symmetry of the wave--guide. Instead,
modest ac magnetic fields are applied to the system. We consider the simple system of a single, closed
conducting loop of radius $r_{\rm ring}$, formed from a conductor of circular cross--section, radius $r_{\rm
wire}$, immersed in a magnetic field, the amplitude of which varies sinusoidally in time. The ac field is
aligned perpendicular to the plane of the coil and is assumed spatially uniform across the area of the coil.
From Faraday's Law, we find that the time varying B--field induces current in the conducting loop, creating an
induced magnetic field about the loop. The external and induced fields cancel symmetrically in a ring,
Fig.~\ref{AdditionOfFields}, the radius of which varies in time. If the potential varies at a frequency much
greater than the atomic motional frequencies then a single trapping radius
is found by averaging the field over one cycle, \cite{Cornell:PRL:95, Arnold:JPhysB:04}. %Due to the symmetry of the system
As the induced currents are inherently alternating, the potentials obtained will benefit from the suppression of
roughness demonstrated in \cite{Westbrook:PRL:07}. The ring trap forms in the plane of the conducting loop and
thus the zero of the instantaneous B--field travels through this trapping radius during each cycle, which may
result in Majorana spin--flip losses of trapped atoms.

Elliptical integrals must be evaluated to obtain the field from a circular loop \cite{Good:EJP:01}. Initial
investigations examined a full spatial integral across the wire for an uniform current density
\cite{skin-depth}. It was found that modeling the current as passing through an infinitesimal region centered on
the wire yielded trapping parameters that agreed to better than 0.5\% with the full model. For the simulations
in this Report, unless stated otherwise, we examine a loop of radius $r_{\rm ring}=1$~cm formed from gold wire
of circular cross--section, with a radius $r_{\rm wire}=0.5$~mm.

The induced emf, $\mathcal{E}$, in a loop is proportional to the rate of change of the magnetic flux, $\phi$,
through the loop,
\begin{equation}\label{eqn:emf}
\mathcal{E} = - \frac{ {\rm d}\,\phi}{{\rm d}\,t} =  L \frac{ {\rm d}\,I}{{\rm d}\,t}  + I\,R\; .
\end{equation}

\begin{figure}[t]
\includegraphics[width=0.48\textwidth]{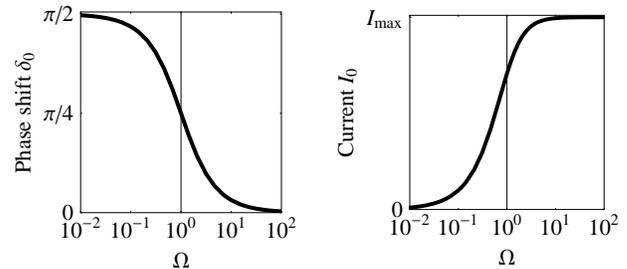}\\
\caption{\label{CurrentPhaseAmplitude} The phase between current in the conducting loop and the driving field
(left) and the induced current (right) as a function of the frequency of the driving field. Note that at low
frequencies the phase induced current lags $\pi$/2 behind the driving field. For the simulations in
Fig.~\ref{TrackingZeros}, $\Omega=4$} \vspace{-0.4cm}
\end{figure}

For a magnetic field of the form $B(t)=B_{\rm dc} + B_0 \, \cos (\omega t)$ the induced current, $I(t)$, may be
expressed as
\begin{equation}\label{eqn:current}
I(t) = \frac{-I_{\rm max}}{\sqrt{1+\Omega^{-2}}}\, \cos(\omega\,t +\delta_0) \,% -{\rm e}^{-\omega t/\Omega}) \,
\end{equation}
where we have introduced the terms $I_{\rm max}=\pi\, r_{\rm ring}^2 \, B_0/L$, $\Omega=\omega\,L/R$ and
$\delta_0=\tan^{-1}(1/\Omega)$. The resistance and inductance of the ring are denoted by $R$ and $L$,
respectively. Eqn.~\ref{eqn:current} applies once turn--on effects, such as overshooting of the current, have
stabilized. The inductance of the loop is dependent on the loop radius, as well as the shape of the conductor
forming the loop \cite{Jackson},
\begin{equation}
\label{eqn:inductance} L \approx \mu_0 \, r_{\rm ring} \, \left( \ln \left( 8 \, r_{\rm ring}\,/\,r_{\rm wire}
\right) - 1.75 \right) \, .
\end{equation}
The resistance is found from the familiar formula incorporating the resistivity, $\rho$, of the conductor,
\begin{equation}
R=2\,\rho \, r_{\rm ring} /  r_{\rm wire}^2\, .
\end{equation}
Due to the resistance and inductance of the conducting loop, the phase and amplitude of the induced current are
dependent on the frequency of the driving field, Fig.~\ref{CurrentPhaseAmplitude}.

\begin{figure}[b]
\includegraphics[width=0.48\textwidth]{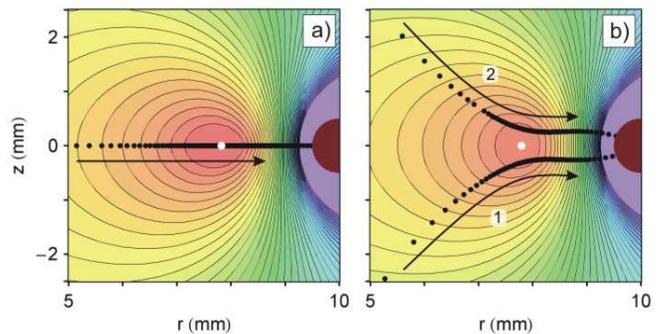}\\
\caption{\label{TrackingZeros} The magnetic potential about the conducting loop showing the effect of the
control parameters. The positions of the time--averaged trap minimum are marked by the white dots, and the
instantaneous zeroes of the B--field by the black dots. The trajectory that the minima follow is indicated by
the numbered, black arrows. For these results $\omega=(2\pi)~30$~kHz, $B_0=100$~G. \textbf{a)}, no additional
fields. \textbf{b)}, a dc quadrupole field, $B_{\rm dcq}=25$~G\,cm$^{-1}$ is added.} \vspace{-0.5cm}
\end{figure}

The trapping potential is obtained by time--averaging over one cycle of the varying fields. In
Fig.~\ref{TrackingZeros}\,\textbf{a)}, a contour plot through the radial and axial plane shows the potential in
the radial plane, with $B_0=100$~G and $\Omega=4$. However, the zeros of the time--varying field pass through
the trapping radius once per cycle, which is undesirable for a magnetic atom trap. If we apply a dc quadrupole
B--field, $B_{\rm dcq}\,\{z\, ,-r/2\}$, centered on the conducting loop we obtain the trapping potential shown
in Fig.~\ref{TrackingZeros}~\textbf{b)}. The instantaneous zeros of the total B--field follow a trajectory that
lies symmetrically about the plane of the ring, removing potential spin--flip losses. Furthermore, the form of
the time--averaged potential is not significantly perturbed by the quadrupole field. The application of a
quadrupole field with axial gradient 25~G\,cm$^{-1}$ forms a stable ring trap 1.76~mm from the surface of the
wire in the plane of the loop, with trapping frequencies of 60~Hz and 49~Hz radially and orthogonal to the plane
of the loop, respectively and a minimum trap depth of 1.2~mK. The average magnetic field at the trap radius is
18.8~G, and the minimum field at that point over one cycle is 9.7~G. As a result of the significant B--field at
all times, the trap's Larmor frequency is typically two orders of magnitude larger than the rate of change of
the magnetic field direction. Trapped atoms will thus adiabatically follow the magnetic field at all times,
further preventing Majorana spin changes of trapped atoms \cite{Ketterle:review:99}.

\begin{figure}[b]
\includegraphics[width=0.45\textwidth]{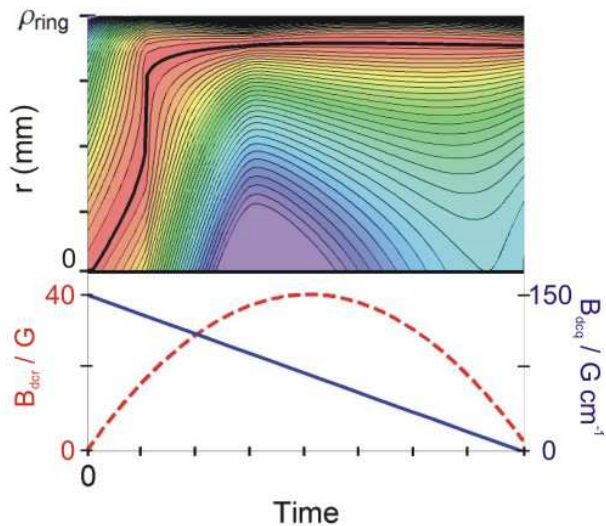}\\
\caption{\label{fig:Quadrupole-to-ring} Demonstrating the transfer from the single quadrupole potential to a
ring trap. The quadrupole field is reduced while a transverse bias field is applied to shift the zero of the
quadrupole toward the ring radius. The top figure shows the potential along a radius parallel to the bias field
as the applied field vary in time, shown in the lower figure.} \vspace{-0.5cm}
\end{figure}

The ring trap is conservative and as such must be loaded with cold atomic samples \cite{AdamsRiis:ProgQE:97}.
The use of a magnetic quadrupole field suggests that a mirror--MOT \cite{Haensch:PRL:99} may be formed above the
center of the loop, which may then be relaxed to load the ring trap. Pre--cooled atoms may be transferred into
the ring trap by spatially overlapping a separate atom trap, such as an optical tweezers
\cite{Pritchard:PRL:02}. Mode--matching of these potentials is likely to prove difficult and instead we propose
a method of deforming the ring potential to provide a 3D magnetic trap. For large B--field gradients the
quadrupole field is the dominant term, forming a trap at the center of the ring. A bias B--field parallel to the
ac driving field shifts the central quadratic trap out of the plane of the loop, providing improved optical axis
for loading the potential with pre--cooled atoms. An additional bias field, $B_{\rm dcr}$, orthogonal to the
first and in the plane of the ring shifts the quadrupole trap towards the ring trap position. To localize atoms
along the ring, and to minimize losses during the transfer, the quadrupole field is reduced and a bias field is
gradually applied. The adiabatic nature of the transfer may be seen in Fig.~\ref{fig:Quadrupole-to-ring}
together with the applied fields. The weak bias B--field along the plane of the loop, has the effect of tilting
the ring and breaks the symmetry of the ring and may be used to localize atoms within the ring
\cite{StamperKurn:PRL:05}.

\begin{figure}
\includegraphics[width=0.45\textwidth]{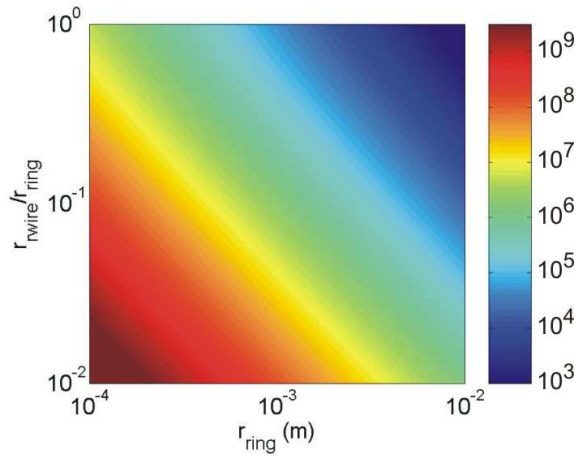}\\ %present_omega0_for_scaling_log_cbar_pcolor.eps}\\
\caption{\label{ScalingR-L} The scaling of drive frequency, $\omega_{\rm drive}\sim\omega=R/L$ as a function of
$r_{\rm ring}$ and $r_{\rm wire}$. The \textit{y}--axis is presented as the ratio of the length scales in the
range of experimental interest.} \vspace{-0.5cm}
\end{figure}

A maximum current density of $j=9300$~A\,cm$^{-2}$ is found for the parameters discussed, with corresponding
Ohmic losses of 5.2~W. The heating losses in the ring scale as $(B_0)^2$, whereas the trap frequencies scale as
$\sqrt{B_0}$, allowing a significant reduction in the power dissipated, if required. These values offer promise
for scaling the ring down, with the consideration of some additional parameters. Firstly, a smaller loop
necessarily has a smaller area, and thus couples less magnetic flux for the same B--field. Secondly, while the
resistance varies as $r_{\rm ring} / r_{\rm wire}^2$, the inductance has a more complex scaling,
Eqn.~\ref{eqn:inductance}. As an example, for a 2~mm diameter gold ring formed of wire with a diameter of
100~$\mu$m, in a 10~G field at 1~MHz, we predict trapping frequencies of (113,\,210)~Hz. Notably, the current
density in this case is the same as for the larger ring, but the total dissipated power is three orders of
magnitude lower. These parameters may be examined in conjunction by considering the frequency scaling $\omega_0
= R/L$, which gives an order of magnitude frequency for the external ac field required to drive the system.
Shown in Fig.~\ref{ScalingR-L}, for ring radii below $\sim100~\mu$m frequencies on the order of 100~MHz are
required to drive the system. At the atom chip scale, the required quadrupole fields can be applied locally
\cite{Pfau:EurPhysJD:03}, allowing for parallelism of the device construction.

A further geometry that is considered is the extension to two concentric, coplanar conducting loops. The induced
currents through each loop are related through their mutual inductance and Eqn.~\ref{eqn:current} is extended to
two coupled equations for these currents. For this system, the frequency of the external B--field should be
chosen $\omega < R/L$ so that the separate currents in the coils run out of phase with the external field. In
the limit of zero drive frequency the currents through each coil are approximately in phase and reach a peak
value when the external field is at a minimum, forming a tight trapping potential between the loops. About the
maximum of the external B--field the current in the loops will have an infinitesimal positive (negative) current
amplitude and the trapping radius forms at the outer (inner) ring. Averaging over a cycle of the ac field the
time--averaged trapping potential forms a ring at approximately the mean radius of the loops. Such a ring
waveguide can have large trapping frequencies due to the proximity of the trapping radius to both wires.

RF--dressed potentials for atoms traps have been a topic of some interest since a proposal by Zobay and Garraway
in 2001 \cite{Garraway:PRL:01}. In this Report we have deliberately chosen parameters to remain in a regime
where the atomic magnetic moment adiabatically follows the total magnetic field and remains in an un--dressed,
$m_F$ state. In this parameter space the modulation frequency of the B--field is much less than the Larmor
frequency and both are significantly larger than the spatial trapping frequencies, as is the case in the TOP
trap. The extension to rf--dressed potentials is readily achievable, albeit requiring higher drive frequencies
than those presented here.

Many applications present themselves for the ring traps proposed here. A circular waveguide is ideal for use in
Sagnac interferometry \cite{Arnold:PRA:06}, the study of sonic black holes \cite{Zoller:PRL:00}, superfluid
circulation in quantum fluids \cite{McCann:PRA:03,Phillips:PRL:07} and soliton propagation
\cite{Kavoulakis:xxx:07,Parker:xxx:07}.

We have proposed and numerically investigated magnetic ring traps that required no external wiring. The scheme
circumvents problems with roughness of the trapping potential by creating a time averaged field that is
inherently smooth with respect to deviations of the current in the conductors. We have identified a mechanism
for eliminating Majorana losses by applying a quadrupole B--field centered on the ring. Large radius $\sim$1~cm
traps have been described that will allow for sensitive interferometric tests, due to the macroscopic area of
the rings. Smaller traps on the micron scale are also proposed which are amenable to micro--fabrication and use
in atom--chip devices.

The authors thank the University of Strathclyde for financial support and Alastair Sinclair, Ifan Hughes and
Charles Adams for discussions.


\begin{references}
\bibitem{skin-depth} At a frequency of 30~kHz the skin depth of gold is $\delta=450~\mu\mathrm{m}\sim r_{\rm wire}$
\end{references}
\end{document}